\documentclass{article}
\usepackage{titlesec}
\usepackage[utf8]{inputenc}
\usepackage{graphicx} 
\usepackage{tikz}
\usepackage{amsmath}
\usepackage{amssymb}
\usepackage{authblk}
\usepackage{biblatex}
\usepackage{subfigure}
\usepackage{placeins}
\usepackage{float}
\usepackage[colorlinks=true, urlcolor=blue, linkcolor=blue, citecolor=red]{hyperref}
\usepackage{graphicx}
\usepackage[margin=1in]{geometry}
\usepackage{fancyhdr}
\title{TSS Graphs for Hadamard Matrices: Real vs Complex}
\author[1]{Wesley Lewis}
\author[1]{Darsh Pareek}

\author[1,2]{Ravi Janjam}
\affil[1]{Researcher}
\affil[2]{Principal Investigator / Chief Scientist }
\affil[1]{\href{https://www.numerikal-labs.org/}{Numerikal Labs}, 1942 Broadway St, Suite 314C, Boulder, CO 80302}
\begin{document}
\maketitle
\begin{abstract}
We have observed that Hadamard matrices whether real or complex lead to a dense output of identical probabilities for any given single input state. Complex Hadamard matrices have unusual orders and introduce phase which prove useful when controlled phase transformation gates are used. By taking a superposed combinations of input states, we discovered that the Hadamards generate non-uniform probabilities which is practically significant towards amplitude amplification without the need for manual parameterization like Grover's operator~\cite{grover1996}. A variety of graph theoretic properties are applied to TSS graphs and their trends were explored. Additionally, graphs turn out to be nearly isomorphic for the same set of input states which has potential applications for developing Quantum Algorithms. 

\end{abstract}

\newpage

\section{Introduction}

Hadamard matrices, first introduced by Jacques Hadamard\cite{hadamard1893resolution} in 1893 in the context of maximal determinants, form a cornerstone of modern discrete mathematics, algebraic combinatorics, and information theory. Formally, a Hadamard matrix $H$ of order $n$ is a square matrix whose entries belong to the set $\{+1, -1\}$ satisfying the orthogonality condition:$$H H^T = n I_n$$where $H^T$ denotes the transpose of $H$ and $I_n$ represents the $n \times n$ identity matrix. Geometrically, the rows (and columns) of $H$ represent mutually orthogonal vectors in $\mathbb{R}^n$, achieving the theoretical upper bound on determinant magnitude for matrices with entries bounded by $\vert{}h_{ij}\vert{} \le 1$, known as Hadamard's Determinant Bound.A necessary condition for the existence of a Hadamard matrix of order $n$ is that $n = 1, 2,$ or $n \equiv 0 \pmod 4$. The long-standing Hadamard Conjecture posits that this necessary condition is also sufficient—namely, that a Hadamard matrix exists for every positive integer multiple of 4. While constructive methods such as Sylvester's Kronecker-product technique, Paley's quadratic residue constructions, and Williamson array-based approaches have proven existence for vast families of orders, finding general construction methods for arbitrary multiples of 4 remains an active open pursuit.

Beyond their theoretical allure, Hadamard matrices have substantial practical utility across diverse engineering and computing domains:
\begin{itemize}
\item \textbf{Error-Correcting Codes}: Forming the mathematical backbone of Hadamard and Reed-Muller codes\cite{1057465}\cite{5745930}, crucial for deep-space telecommunications.
\item \textbf{Signal Processing and Transforms}: Powering the Fast Walsh-Hadamard Transform (FWHT), widely used in image compression\cite{897063}, spectral analysis\cite{Mende:93}, and CDMA wireless communication systems.
\item \textbf{Quantum Computing}: Constructing quantum gates (e.g., the Walsh-Hadamard transform creating equal superpositions) and quantum error-mitigation schemes\cite{windisch2024hadamard}.
\item \textbf{Cryptography and Optimization}: Supplying orthogonal sequences and evaluating Boolean functions via the Walsh--Hadamard transform, linking the structural invariants and permanents of Sylvester Hadamard matrices to maximal non-linearity (e.g., bent functions) to resist linear cryptanalysis\cite{math9020177}
\end{itemize}

As we get equal superpositions with Hadamard, we perform first order and second order statistics to find interesting patterns, trends, gaps and convergence of states where we get number of inputs \textgreater  number of outputs.
In this paper, we discuss our findings while using Hadamard matrices as gates in generating TSS graphs\cite{lewis2026}.

Hadamard matrices containing all 1's in the first row and first column is a normalized hadamard matrix. Any matrix of the same order which can be represented by switching the rows and columns of the matrix and get a normalized matrix is equivalent to the normalized matrix. Any third matrix which also can be normalized is considered equivalent to the third matrix. 

A \textit{complex Hadamard matrix} $C$ of order $c$ is a matrix all of whose elements
are $+1$, $-1$, $+i$ or $-i$ and which satisfies $CC^* = cI_c$, where $i = \sqrt{-1}$.
It is conjectured that a complex Hadamard matrix exists for every even $c$.
\subsection{Classical Graph Theoretic Analysis on TSS Graphs}
To formally interpret the computational transformations of real and complex Hadamard operators, we model their input--output state mappings through graph theory under the Topological Structure of Superpositions (TSS) framework \cite{lewis2026}. Formally, a \textbf{directed graph} (or digraph) is an ordered pair $G = (V, E)$, where $V$ represents a finite set of \textbf{vertices} and $E \subseteq V \times V$ denotes a set of ordered pairs called directed \textbf{edges}  \cite{diestel2012graph}. Within our computational setting, vertices $v \in V$ correspond to addressable computational basis states, while directed edges $(u, v) \in E$ signify admissible computational transitions from an input basis state $u$ to an output basis state $v$ driven by constructive and destructive interference \cite{aharonov2001}. A directed edge of the form $(u, u) \in E$ constitutes a \textbf{self-loop}, indicating that a computational state is invariant under the local transformation. By abstracting unitary operators into discrete combinatorial structures rather than dense transition matrices, TSS graphs provide a direct topological visualization of the global computational structure \cite{lewis2026, tadej2006concise}.

A central challenge in evaluating these networks is distinguishing genuinely distinct computational geometries from trivial relabelings of the underlying basis states. We address this through formal graph-theoretic metrics and equivalence relations:

\begin{itemize}
    \item \textbf{Graph Isomorphism and Invariants:} Two directed graphs $G_1 = (V_1, E_1)$ and $G_2 = (V_2, E_2)$ are \textbf{isomorphic} (denoted $G_1 \cong G_2$) if there exists a bijection $f: V_1 \to V_2$ such that $(u, v) \in E_1 \iff (f(u), f(v)) \in E_2$ \cite{diestel2012graph}. Isomorphic graphs preserve edge-connectivity and network topology across relabelings. Although all evaluated operators share the foundational Hadamard orthogonality property, their generated TSS graphs partition into uniquely sized \textbf{isomorphism classes}. To uniquely distinguish and systematically order these equivalence classes without exhaustive graph-matching searches, we evaluate degree polynomials as complete graph invariants \cite{lewis2026}.
    
    \item \textbf{Strongly Connected Components (SCCs):} A directed graph is \textbf{strongly connected} if every vertex is reachable from every other vertex via a directed path \cite{diestel2012graph}. The maximal strongly connected subgraphs partition $V$ into disjoint \textbf{strongly connected components (SCCs)} \cite{tarjan1972depth}. In quantum information routing, the number of SCCs ($|SCC|$) measures the degree of structural fragmentation or state clustering: $|SCC| = 1$ reflects complete reciprocal reachability across the Hilbert space, whereas higher SCC counts signify isolated or strictly feed-forward state manifolds.
    
    \item \textbf{Simple Directed Cycles:} A \textbf{simple cycle} is defined as a closed directed walk $(v_0, v_1, \dots, v_k = v_0)$ containing no repeated vertices other than the start and end node ($v_i \neq v_j$ for all $0 \le i < j < k$) \cite{johnson1975finding}. Within TSS graphs, simple cycles delineate localized, closed-loop feedback pathways where the operator enables recurring state preservation rather than purely dissipative transitions \cite{childs2009}.
    
    \item \textbf{Fundamental Combinatorial Metrics:} Low-level structural properties---specifically vertex cardinality $|V|$, total edge count $|E|$, and self-loop frequencies---quantify the baseline sparsity, edge densification, and immediate state span induced by varying matrix dimensions and input superposition degrees \cite{lewis2026, tadej2006concise}.
\end{itemize}

These distinct topologies demonstrate that varied superposition inputs collapse into a constrained, highly symmetric family of routing networks \cite{lewis2026}. We analyze these graph-theoretic invariants, their parametric scaling, and their algorithmic implications comprehensively in Sections \ref{sec:results} and \ref{sec:discussion}.

\subsection{Matrix analysis observation}

\begin{table}[!ht]
    \centering
    \begin{tabular}{|l|l|p{2cm}|p{3cm}|p{4cm}|p{4cm}|}
    \hline
        \textbf{Matrix ID} & \textbf{Size} & \textbf{Input} & \textbf{Output} & \textbf{pp\_re} & \textbf{pp\_im} \\ \hline
        M002 & 8 & (3, 0) & (5, 4, 3, 1, 0) & (-0.47, 0.16, 0.23, 0.47, -0.39) & (0, 0.44, 0.33, 0, 0.11) \\ \hline
        M004 & 8 &  (3, 0) & (7, 6, 5, 4, 3, 2, 1, 0) & (0.44, 0.06, -0.44, 0.25, -0.06, 0.25, -0.25, 0.25) & (-0.17, 0.21, 0.17, -0.26, -0.21, -0.2, 0.26, -0.2) \\ \hline
        M005 & 8 &  (3, 0) & (6, 5, 4, 3, 2, 1, 0) & (0.18, 0.25, 0.18, 0, -0.18, -0.25, -0.18) & (-0.43, 0.25, -0.07, 0.5, -0.07, 0.25, 0.43)  \\ \hline
        M005 & 8 &  (4, 0) & (6, 4, 2, 0) & (0.35, 0.35, -0.35, -0.35) & (-0.35, 0.35, 0.35, -0.35) \\ \hline
        M006 & 8 &  (5, 1) & (5, 4, 3, 2, 1) & (-0.35, 0.44, -0.35, -0.18, 0.44) & (0, 0.23, 0, -0.47, 0.23)  \\ \hline
        M006 & 8 &  (6) & (7, 6, 5, 4, 3, 2, 1) & (0, -0.25, -0.25, 0.37, 0, 0.37, -0.35) & (0.38, -0.28, -0.28, 0.05, 0.38, 0.05, -0.28) \\ \hline
        MS002 & 16 & (10, 9, 8, 7, 6, 5, 4, 3, 2, 1) & (15, 11, 10, 8, 4, 3, 1, 0) & (0, 0, 0, 0, 0, 0, 0, 0) & (0.35, -0.35, 0.35,0.35, -0.35, 0.35, 0.35, 0.35) \\ \hline
    \end{tabular}
     \caption{Comparison between state transitions: input state (equiprobable superpositions), output state, pp\_re: Re(p.f) for real part and pp\_im: Im(p.f) for imaginary part of probability factor (p.f)}
    \label{tab:comparison_between_state_transitions}
\end{table}

In table 1 \ref{tab:comparison_between_state_transitions} we see a small sample of the possible results recorded during our analysis. Here we have separated the real and imaginary part of the p.f. in pp\_re for real part and pp\_im for imaginary part respectively. At a glance the table is not easily  understandable, thus we go perform a second order statistical analysis to gain perspective. The aim of the second order analysis was to understand cases where there were converging states i.e. no. of input states \> no. of output states. As for hadamard matrices the opposite is observable to a large extent. 

In table \ref{tab:comparison_interference_data} we see the total number of convergence cases recorded in the analysis of these hadamard matrices ordered by Matrix ID. Combining these 2 tables we get a understanding that the no. of convergence is not related to the order of the matrix but rather in the differences of placement of elements of each matrix. MS002 is a order 16 matrix (with padding of 4). We find the maximum number of convergence cases in MS002. This could be to also because of the entanglement space in terms of binary representation is higher that order 8 matrices and since we've taken only a sample of the entangled state we record such a high number. 

\begin{table}[!ht]
    \centering
    \begin{tabular}{|l|l|l|l|l|l|l|}
    \hline
\textbf{Matrix ID}  & \textbf{Convergence} &  \textbf{permutations} & \textbf{complex p.f} &  \textbf{real p.f} & \textbf{min states}&\textbf{ max states} \\ \hline
        M002  & 688 &  4 & 0 & 2 & 5 & 8 \\ \hline
        M004  & 120 & 4 & 0 & 0 & 8 & 8 \\ \hline
        M005  & 128 & 4 & 1 & 0 & 7 & 7 \\ \hline
        M005  & 128 &  4 & 0 & 0 & 4 & 4 \\ \hline
        M006  & 216 &  4 &  0 & 2 & 5 & 7 \\ \hline
        M006  & 312 &  2 &  2 & 0 & 7 & 7 \\ \hline
        MS002 & 27448 & 1024 & 8 & 0 & 8 & 14 \\ \hline
    \end{tabular}
    \caption{Comparision between possible interference permutations, purely real and complex probability factor (p.f), min no. of states recorded, max no. of states recorded and convergence cases}
    \label{tab:comparison_interference_data}
\end{table}

Please note that for MS002 we have taken a sample of 40000 permutations of an entangled state as the total number of iterations increases for order 16 matrices with a very large computation time. Convergence is the total number of convergence cases recorded in the entire analysis for MS002. Permutations is the observed number of possibilities for constructive and destructive interference for that particular input state.

\section{Methodology}
We have sourced the real and complex Hadamard matrices from 2 sources. 
1) Neil Sloane's library of hadamard matrices\cite{SloaneHadamard} which contains all hadamard matrices of orders n up through 28, atleast one of every order n up through 256. 
2)  Catalog of complex hadamard matrices \cite{CHM_Catalogue}

We use the following steps in constructing the imperfect graphs from Hadamard matrices.
1) Apply all states one by one to the gate and get a set of output gates
2) If the state is an entangled state then we break them into a power 2 gate and then combine them.
3) While combining we have the possibility for constructive and destructive interference. 
4) This gives us more possibilities for just a single state.
5) Construct a graph without the probabilities

Combining all the states graphs together doesn't give us anything interesting. It's either a complete graph or a near complete graph. 
\section{Results}

\subsection{Hardware used for analysis}
We have used 8 core processor for the processing this data with a processing time of on average 5 mins.\\ 
We have a total of 15 matrices. All of these matrices are $<$ order 16. We have taken a mixture of real hadamard and complex hadamard matrices. For matrices with 8 $<$ order $<$ 16 we have taken the entangled set of qubits for our analysis. We take combinations of (order/2) 1's and (order/2) 0's for creating the entangled set. For order 16 with 8 1's and 8 0's in the sequence, we get around 75000 such combinations. Out of these combinations, we take a sample of 4000 for generating the plots and 10000 for generating the statistics. 

\subsection{Frequency observations}
For input qubit of 2\^n we get a common graph of 2\^n going to all other possible nodes. If we have order 8 matrix. Then for input qubit 8 we'll have the $E = \{(u, i) \mid i \in \{0, 1, \dots, n\} \}$ 

This behavior is recorded across real and complex hadamard matrices. 

\begin{figure}[ht]
    \centering
    \includegraphics[width=1\linewidth]{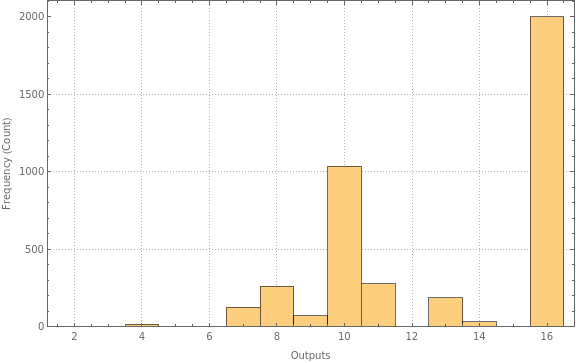}
    \caption{Frequency plot of outputs in MS002}
    \label{fig:frequency_plot_MS002}
\end{figure}

For the above plot, we take entangled states of a sequence containing equal number of 0's and 1's. If we have a n-bit sequence, n/2 bits are 0's and n/2 bits are 1's. We see a pattern where there are a few gaps in the outputs. The plot also contains some local maxima for the 2\^n values. 

\begin{figure}
    \centering
    \includegraphics[width=1\linewidth]{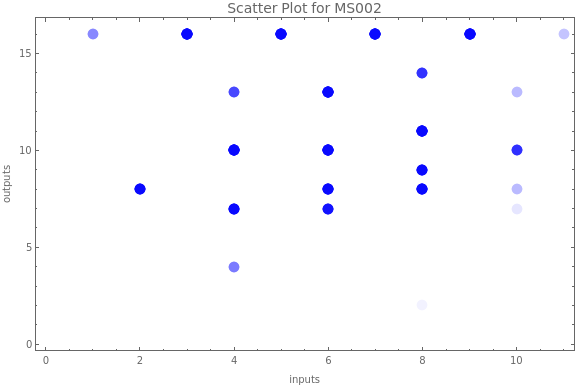}
    \caption{Scatter plot for no. of inputs vs no. of outputs}
    \label{fig:input_output_relation}
\end{figure}

The most notable thing we find here is that the gate MS002 has a convergence of states. It is is very light in shade to represent the frequency of this occurrence. Although rare, this is a notable insight that we can use to construct circuits where we have the need to reduce the number of states for that particular value. 

\begin{figure}
    \centering
    \includegraphics[width=1\linewidth]{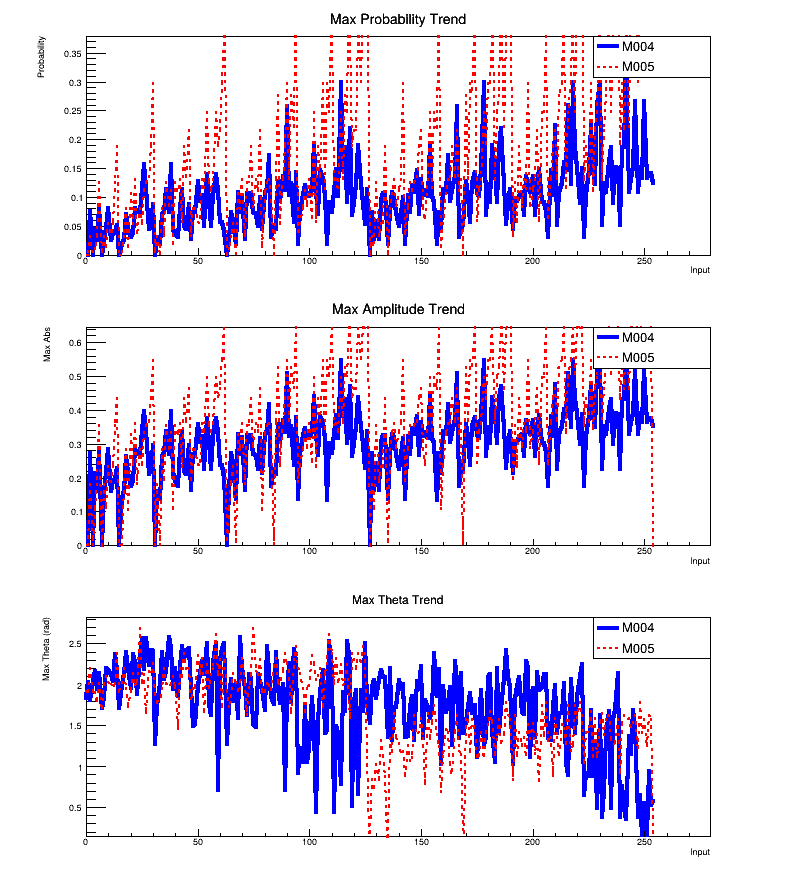}
    \caption{Trend comparison between M004, M005}
    \label{fig:trend_M0004_M005}
\end{figure}

Of all the possible combinations we have processed, the maximum difference between the trends is found between M004 and M005. These are complex hadamard matrices of order 8. M004 has higher probability factor values in comaprison with M005. This is because the number of output states recorded for M004 is less than M005. Thus, the p.f values are higher. 

\label{sec:results}

\subsection{Graph Invariants and Isomorphism Classes}
Applying the construction methodology outlined in \cite{lewis2026} to the evaluated Hadamard matrices generated over 4,500 directed graphs, with individual matrices yielding between 15 and 2,800 graphs. To eliminate redundancy, the graphs derived from each matrix were partitioned into distinct isomorphism classes. Across all evaluated matrices, the number of isomorphism classes ranged from a minimum of 6 to a maximum of 97. 

Because the degree polynomials functioned as complete invariants for all generated graphs in our dataset, they were utilized to uniquely distinguish and sort the isomorphism classes in ascending order. Evaluated structural properties of these graphs exhibited a monotonic increase with respect to this polynomial ordering.
\subsubsection{Fundamental Graph Invariants}
We evaluated the variation of low-level topological invariants across the ordered isomorphism classes, specifically focusing on total edge count and the frequency of self-loops.

\begin{figure}[H]
    \centering
    \includegraphics[width=0.95\linewidth]{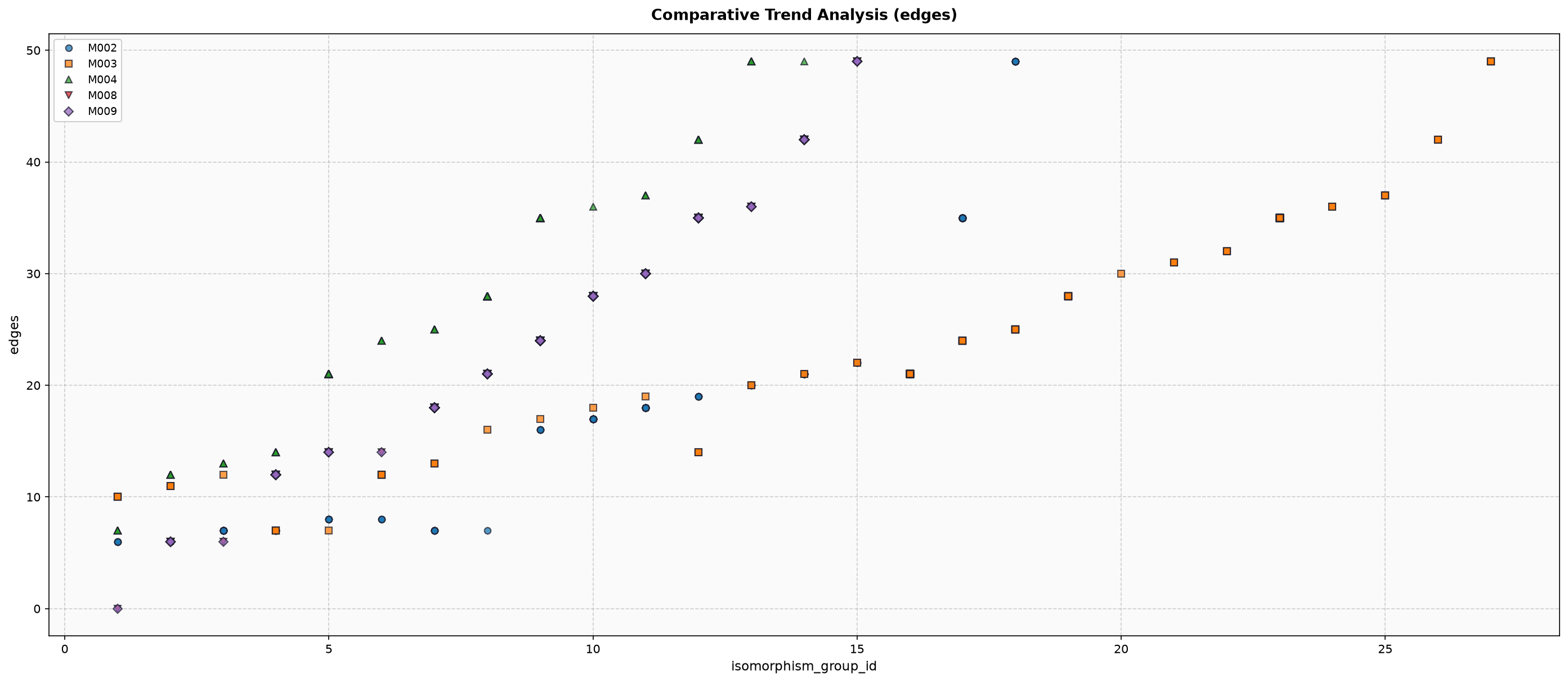}
    \caption{Total edge count as a function of the isomorphism class index (ordered by degree polynomial) across matrices $M002$, $M003$, $M004$, $M008$, and $M009$.}
    \label{fig:edge_trends}
\end{figure}

Figure~\ref{fig:edge_trends} illustrates the relationship between the isomorphism class index (ordered by degree polynomial) and total edge count across representative matrices ($M002$, $M003$, $M004$, and $M009$). Across all evaluated matrices, the edge count exhibits a clear positive trend with respect to the class index. For example, in matrix $M003$, edge counts progress from 10 at class index 1 to a maximum of 49 at index 27, though minor non-monotonic fluctuations occur locally (e.g., at indices 4 and 12). Matrices with fewer isomorphism classes ($M002$, $M004$, $M009$) demonstrate steeper edge accumulation rates over their respective class domains.
\FloatBarrier

\begin{figure}[H]
    \centering
    \includegraphics[width=0.95\linewidth]{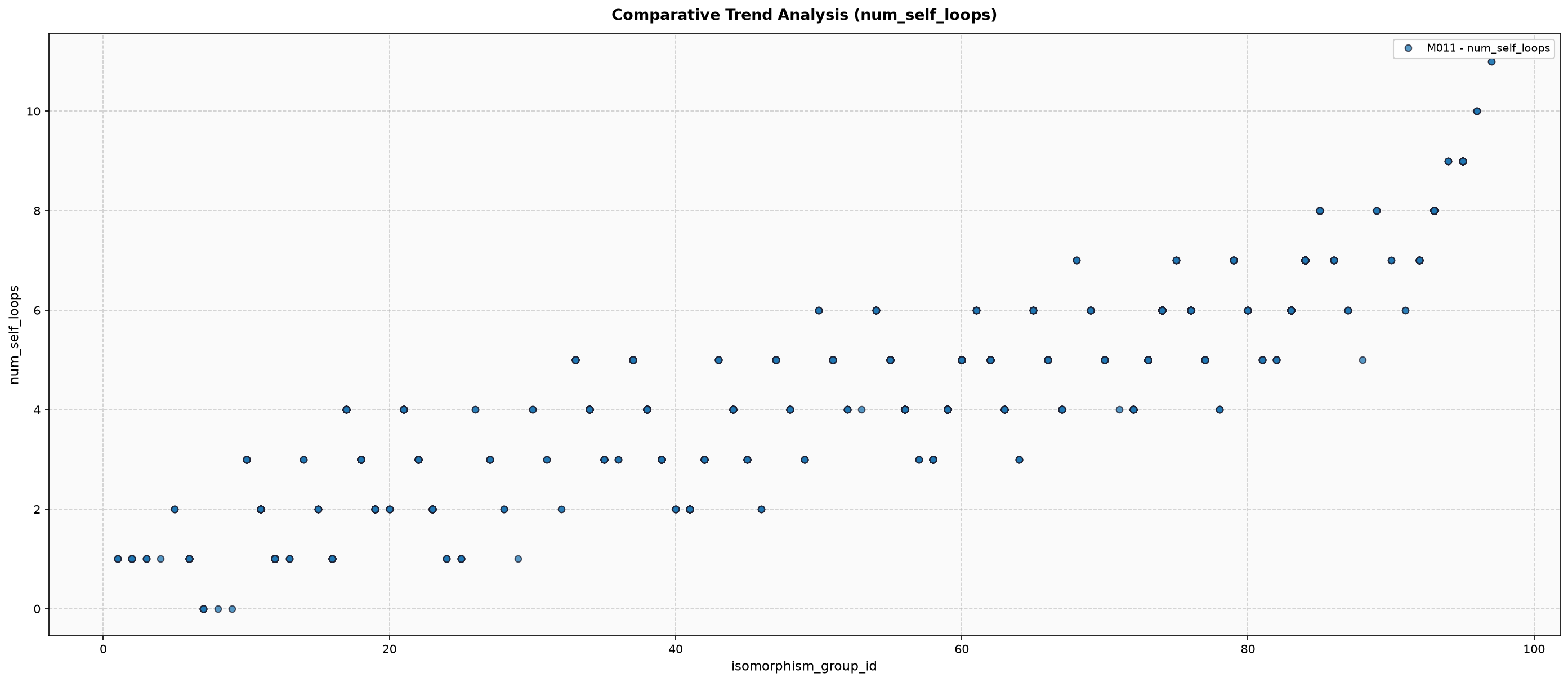}
    \caption{Distribution of self-loops across the 97 isomorphism classes of matrix $M011$, ordered by degree polynomial.}
    \label{fig:self_loops}
\end{figure}

Figure~\ref{fig:self_loops} tracks the distribution of self-loops across the 97 isomorphism classes derived from matrix $M011$. While local oscillations are prevalent across adjacent classes—such as fluctuations between 0 and 4 self-loops within the first 30 classes—the macro-level envelope shifts upward, reaching a plateau around 4--6 self-loops across intermediate classes (indices 35--80) before terminating at a peak of 11 self-loops at class index 97.
\FloatBarrier 
\subsubsection{Higher-Order Topological Properties}
To evaluate macro-scale structural connectivity and symmetry across the graph ensemble, we analyzed the number of strongly connected components (SCCs), simple cycle counts, and isomorphism class cardinalities.

\begin{figure}[H]
    \centering
    \includegraphics[width=0.95\linewidth]{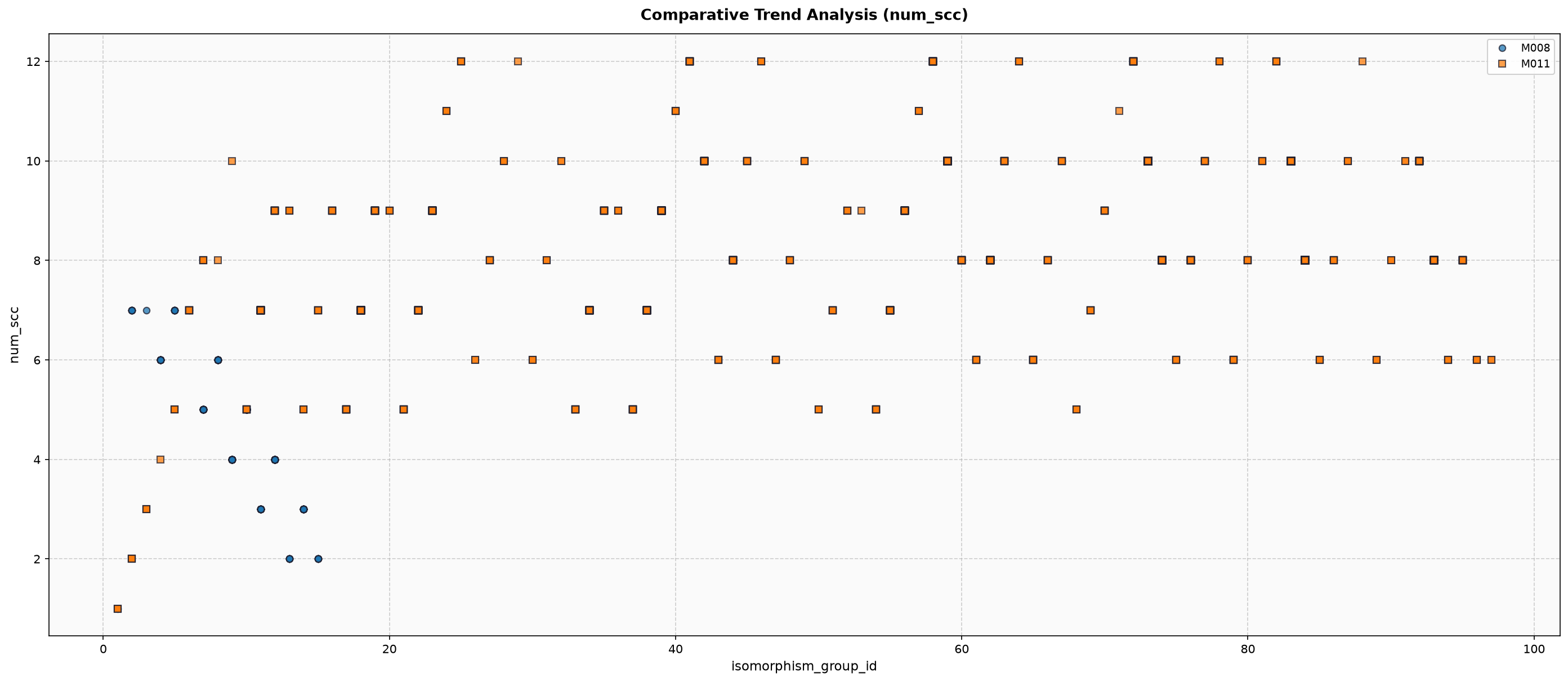}
    \caption{Number of strongly connected components ($|\mathrm{SCC}|$) as a function of the isomorphism class index (ordered by degree polynomial) for matrices $M008$ and $M011$.}
    \label{fig:num_scc}
\end{figure}

\paragraph{Strongly Connected Components}
The progression of SCC counts exhibits contrasting behavior across different matrix families, as shown in Figure~\ref{fig:num_scc}. For matrix $M008$, the number of SCCs decreases monotonically from 7 down to 2 as the isomorphism class index increases across its 15 classes. Conversely, matrix $M011$ begins as a single strongly connected graph ($|\mathrm{SCC}| = 1$ at index 1), rapidly fragments into up to 12 components by index 25, and oscillates between 5 and 12 components across the remaining classes.

\begin{figure}[H]
    \centering
    \includegraphics[width=0.95\linewidth]{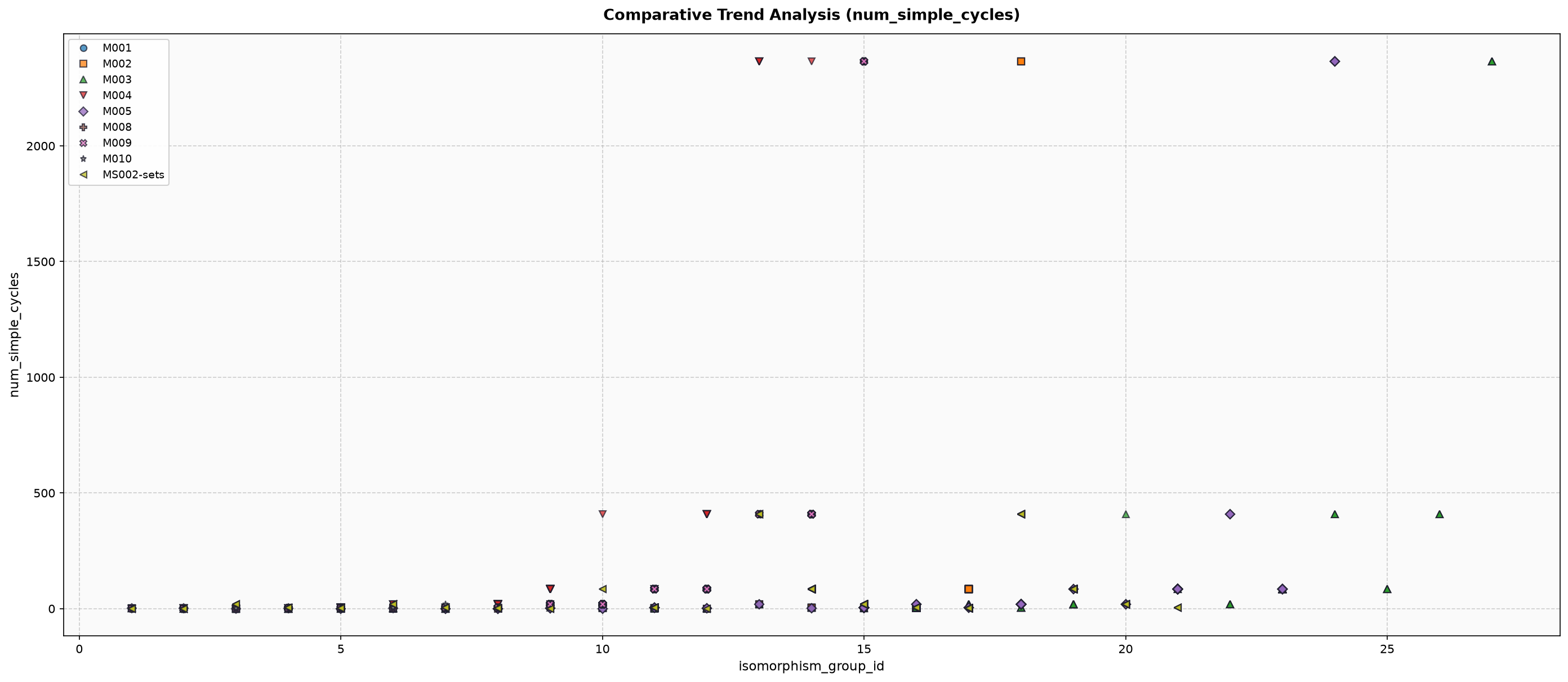}
    \caption{Simple cycle counts across sorted isomorphism classes for representative Hadamard matrices ($M001$--$M004$, $M008$--$M010$, and $MS002$).}
    \label{fig:num_simple_cycles}
\end{figure}

\FloatBarrier
\paragraph{Simple Cycle Distribution}
Figure~\ref{fig:num_simple_cycles} depicts the distribution of simple cycles across isomorphism classes. While lower- and intermediate-indexed classes contain minimal cycle counts ($\le 10$), the metric exhibits discrete, quantized surges at higher class indices. Specifically, several distinct matrices converge to identical cycle counts at their terminal indices, notably reaching discrete plateaus at 84, 412, and an upper bound of 2,365 simple cycles (e.g., observed at index 14 for $M004$, index 15 for $M008$/$M009$, index 17 for $M002$, and index 27 for $M003$). In these instances, graphs exhibiting identical cycle counts across differing degree polynomials correspond to structural graph complements or mutually dual configurations.

\FloatBarrier

\begin{figure}[H]
    \centering
    \includegraphics[width=0.95\linewidth]{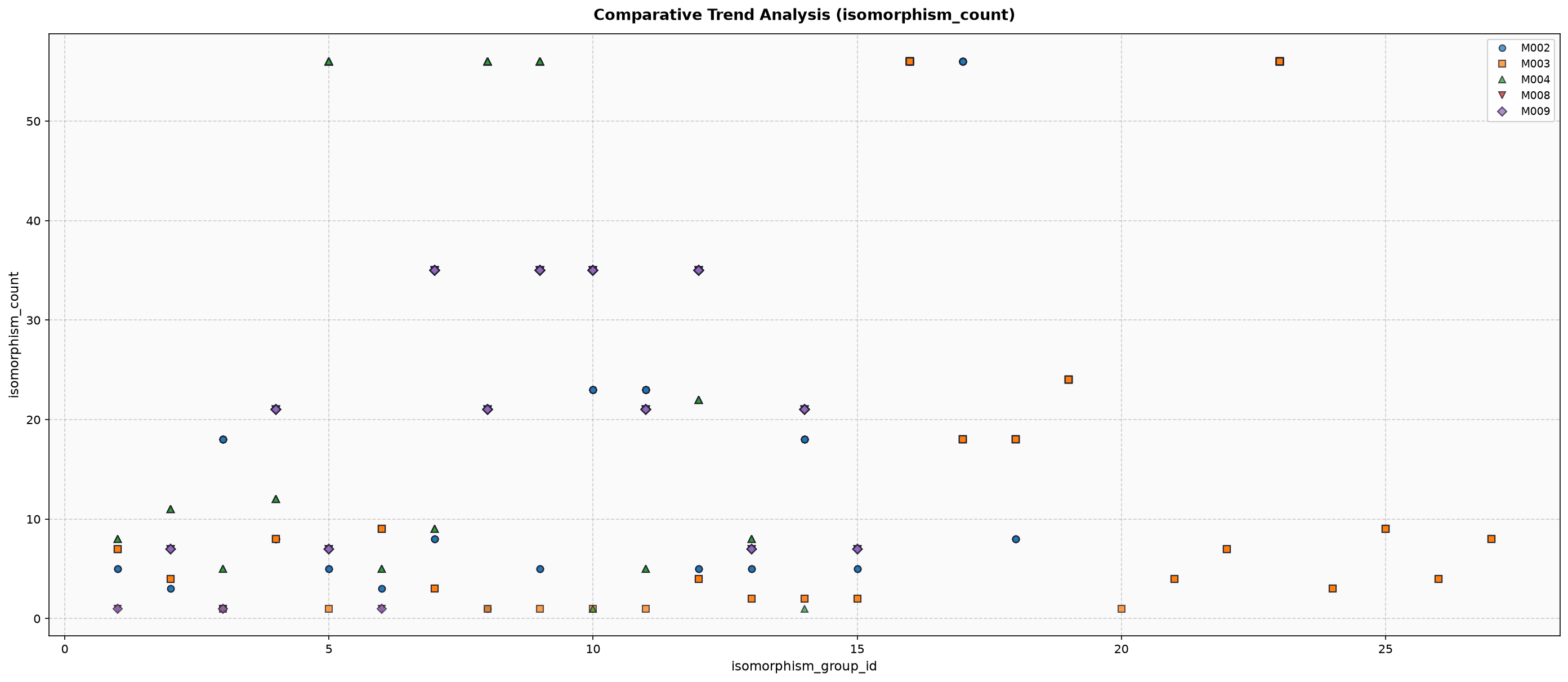}
    \caption{Isomorphism class cardinality (number of constituent isomorphic graphs) across sorted class indices.}
    \label{fig:isomorphism_count}
\end{figure}

\paragraph{Isomorphism Class Cardinality}
The distribution of graphs per isomorphism class is illustrated in Figure~\ref{fig:isomorphism_count}. Across all matrix families, the majority of equivalence classes exhibit low cardinalities (predominantly between 1 and 10 constituent graphs). However, class sizes are non-uniform: specific classes exhibit pronounced degeneracy peaks, clustering at cardinalities of 21, 35, and an upper ceiling of exactly 56 graphs per class across matrices $M002$, $M003$, and $M004$.

\FloatBarrier 
\subsection{Correlation Between Matrix Inputs and Graph Invariants}
Having analyzed matrix structures and the generated graph topologies independently, we now examine the direct mapping between input characteristics and resulting graph invariants. Specifically, we evaluate graph topology as a function of two input parameters: the matrix dimension and the number of superimposed basis states. We first correlate these input parameters with fundamental graph metrics (vertex count, edge count, and self-loop frequency) before evaluating their impact on higher-order topological features.

\subsubsection{Fundamental Graph Properties and Input Scaling}
We next analyzed how input specifications—specifically the matrix dimension ($n_{\mathrm{inputs}}$) and the input state vector sparsity (number of non-zero entries, representing the degree of superposition)—govern primitive graph metrics.

\begin{figure}[H]
    \centering
    \includegraphics[width=0.95\linewidth]{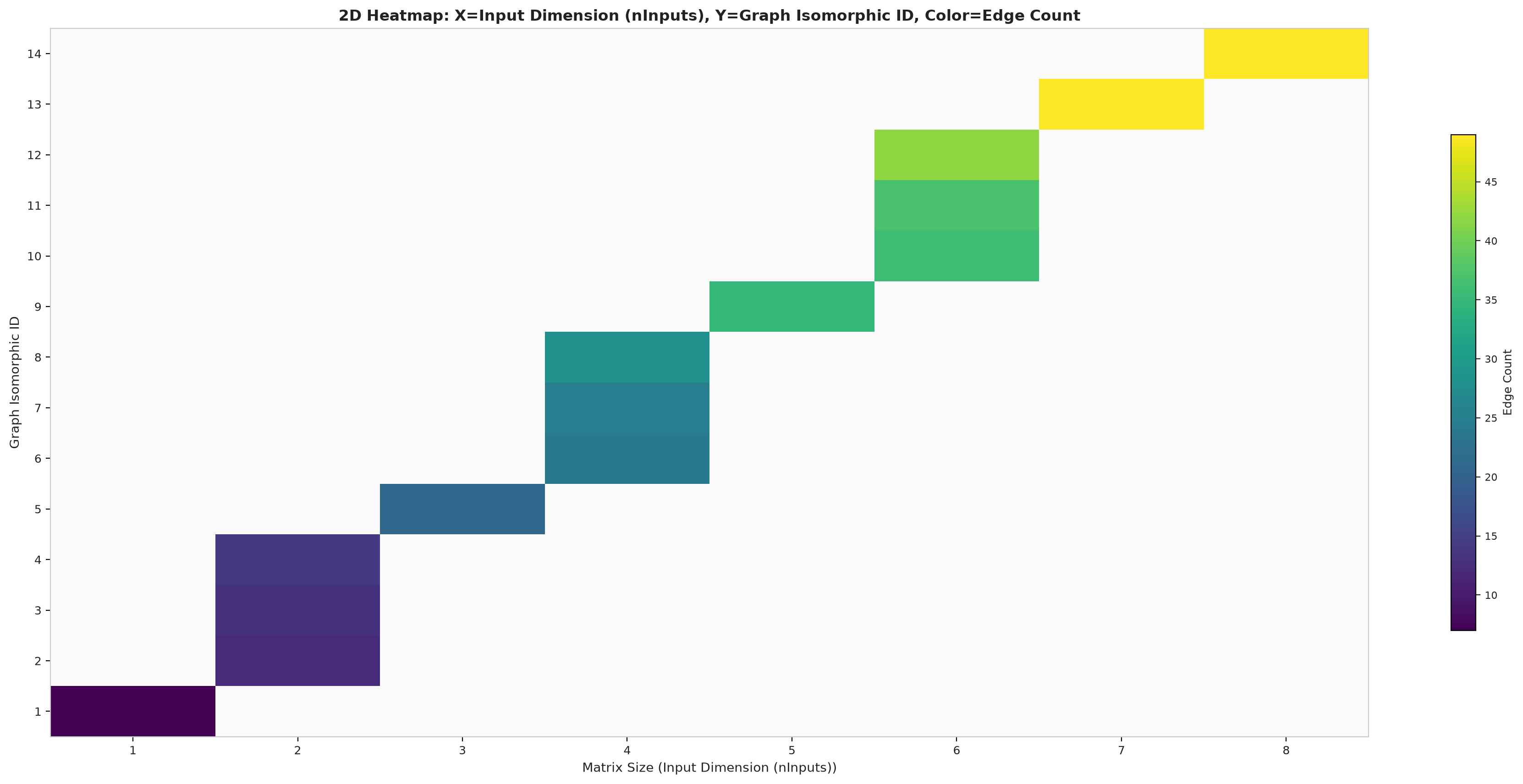}
    \caption{Two-dimensional heatmap showing the correspondence between input matrix dimension ($n_{\mathrm{inputs}}$), resulting isomorphism class index, and edge count for matrix $M004$.}
    \label{fig:heatmap_edgecount}
\end{figure}

\paragraph{Scaling with Matrix Dimension}
Figure~\ref{fig:heatmap_edgecount} depicts the relationship between input matrix dimension, isomorphism class index, and edge count for matrix $M004$. The generation displays a strict block-diagonal partition: each input dimension generates a mutually disjoint subset of isomorphism classes. For example, dimension 1 maps uniquely to class index 1, dimension 2 spans indices 2--4, and dimension 8 maps exclusively to class index 14. Concurrently, the edge count exhibits a monotonic progression along this diagonal, scaling from 7 edges at dimension 1 (dark purple) to an upper ceiling of 49 edges at dimensions 7 and 8 (bright yellow).

\begin{figure}[H]
    \centering
    \includegraphics[width=0.95\linewidth]{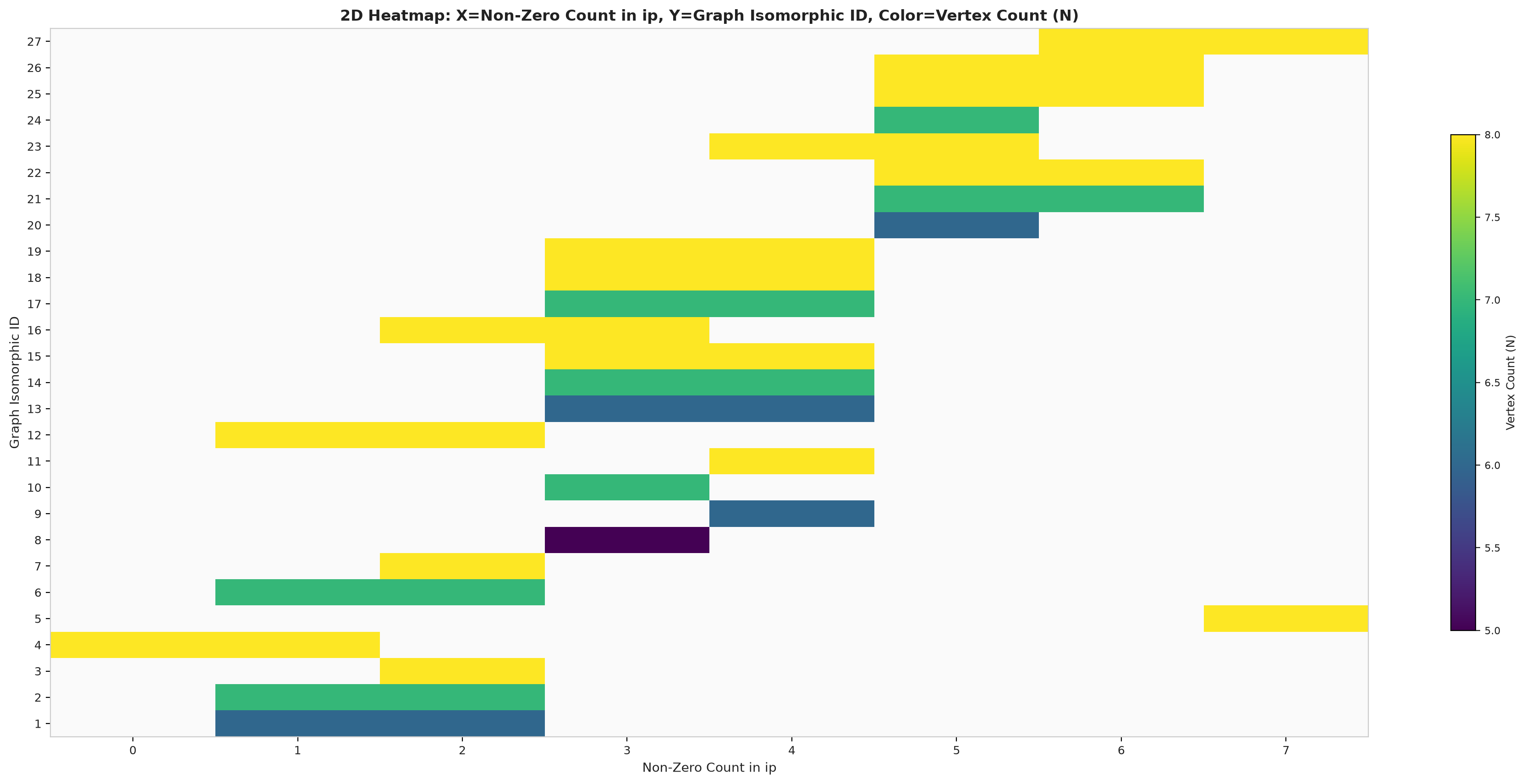}
    \caption{Two-dimensional heatmap illustrating graph vertex count ($N$) as a function of the input vector non-zero count (superposition degree) and isomorphism class index for matrix $M003$.}
    \label{fig:heatmap_vertex_count}
\end{figure}

\paragraph{Scaling with Superposition Cardinality}
Figure~\ref{fig:heatmap_vertex_count} illustrates the effect of input vector support size (number of superimposed non-zero components) on graph vertex count ($N$) across the 27 isomorphism classes generated by matrix $M003$. Vertex counts remain bounded within the interval $5 \le N \le 8$. While lower-order isomorphism classes (indices 1--7) are predominantly generated by sparse vectors containing 1 to 2 non-zero elements, access to higher-order classes (indices 20--27) requires denser superpositions of 5 to 7 non-zero components. Across nearly all superposition levels, the majority of classes saturate at the maximum observed order of $N = 8$ vertices (represented by yellow bands).

\FloatBarrier
\subsubsection{Higher-Order Topological Properties and Input Scaling}
We evaluated the scaling of macro-topological invariants—simple cycle counts, strongly connected components (SCCs), and isomorphism class cardinalities—against input matrix dimensions ($n_{\mathrm{inputs}}$).

\begin{figure}[H]
    \centering
    \includegraphics[width=0.95\linewidth]{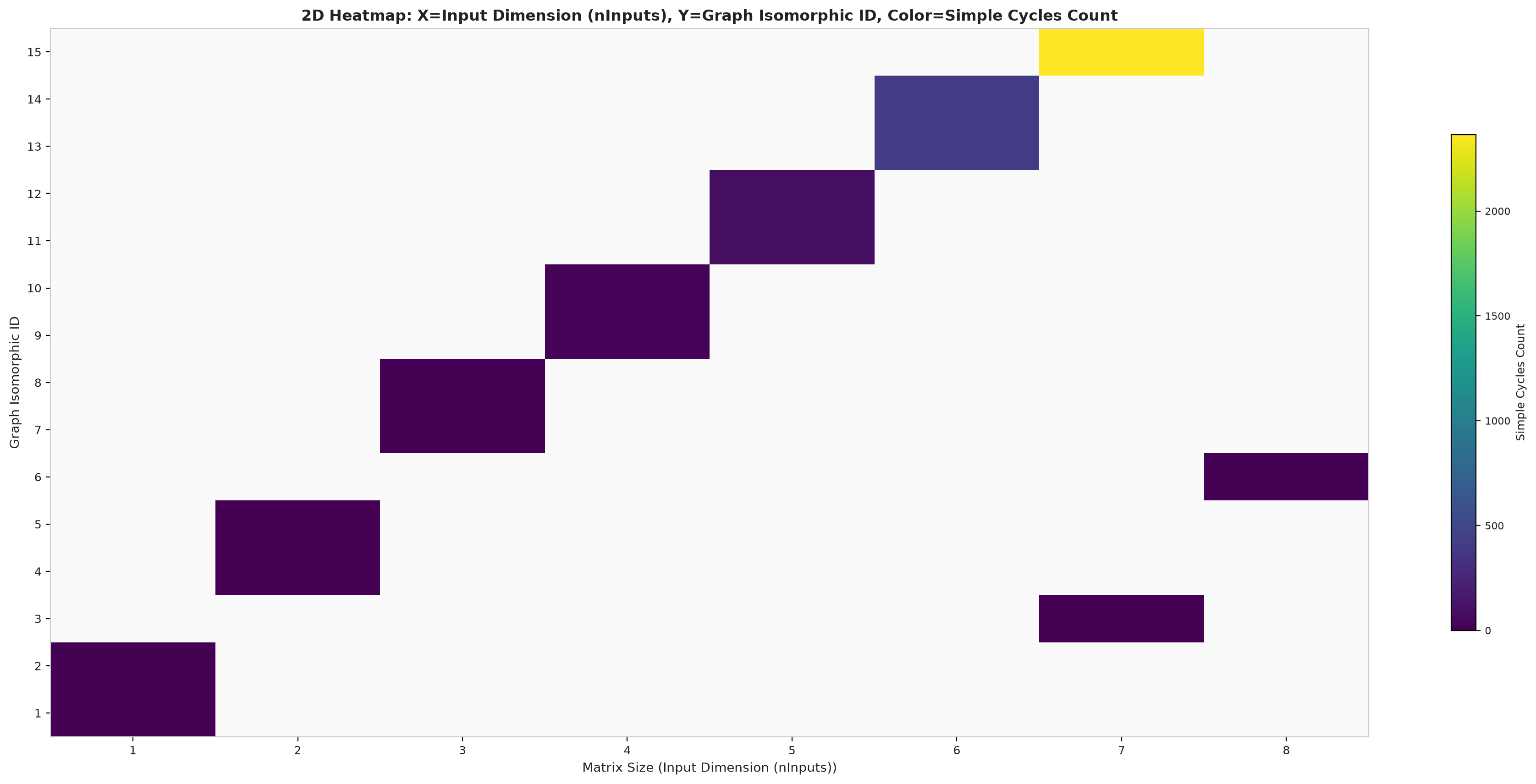}
    \caption{Two-dimensional heatmap correlating input matrix dimension ($n_{\mathrm{inputs}}$), isomorphism class index, and simple cycle count for matrix $M009$.}
    \label{fig:heatmap_simple_cycles}
\end{figure}

\paragraph{Simple Cycle Concentration}
Figure~\ref{fig:heatmap_simple_cycles} displays simple cycle counts across input dimensions and isomorphism classes for matrix $M009$. Cycle formation remains negligible ($\le 10$) across dimensions 1 through 5. A modest increase emerges at dimension 6 (spanning indices 13 and 14), while the maximal cycle count of 2,365 is localized exclusively to dimension 7 at class index 15. The remaining classes generated at dimension 7 (indices 3 and 6) retain near-zero cycle counts, demonstrating that the structural conditions necessary for extensive cycle generation are restricted to specific, high-index classes at this input dimension.

\begin{figure}[H]
    \centering
    \includegraphics[width=0.95\linewidth]{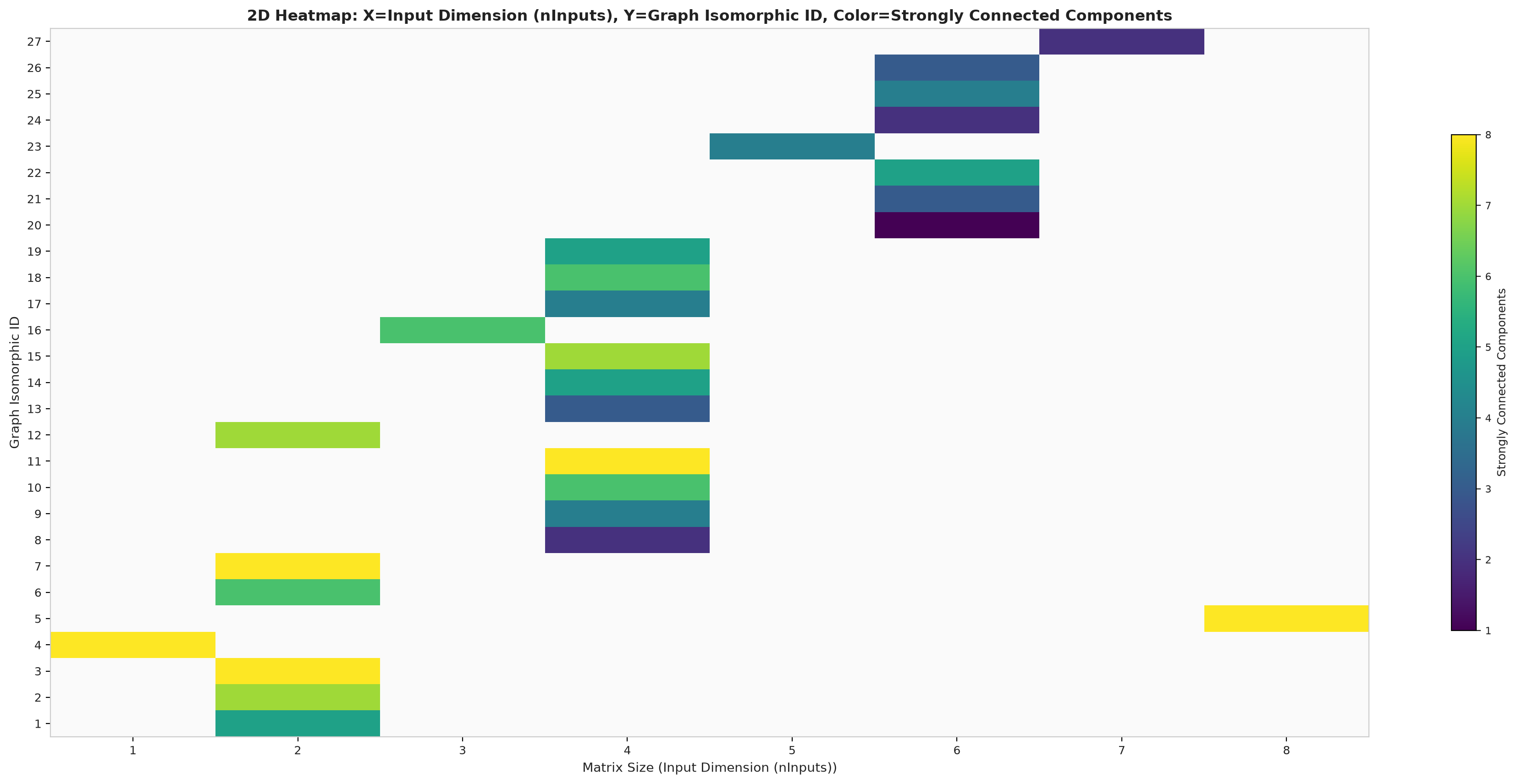}
    \caption{Two-dimensional heatmap of strongly connected component counts ($|\mathrm{SCC}|$) across input matrix dimensions ($n_{\mathrm{inputs}}$) and isomorphism classes for matrix $M003$.}
    \label{fig:heatmap_scc}
\end{figure}

\paragraph{Strongly Connected Component Dynamics}
Figure~\ref{fig:heatmap_scc} examines the distribution of strongly connected components across input dimensions for matrix $M003$. Extremal input dimensions produce heavily partitioned topologies: dimensions 1, 2, and 8 generate graphs with elevated component counts ($6 \le |\mathrm{SCC}| \le 8$), indicating fragmented structural clusters. Conversely, intermediate dimensions (specifically 4 and 6) yield the lowest component counts, with several classes consolidating down to $|\mathrm{SCC}| = 1\text{--}3$.

\begin{figure}[H]
    \centering
    \includegraphics[width=0.95\linewidth]{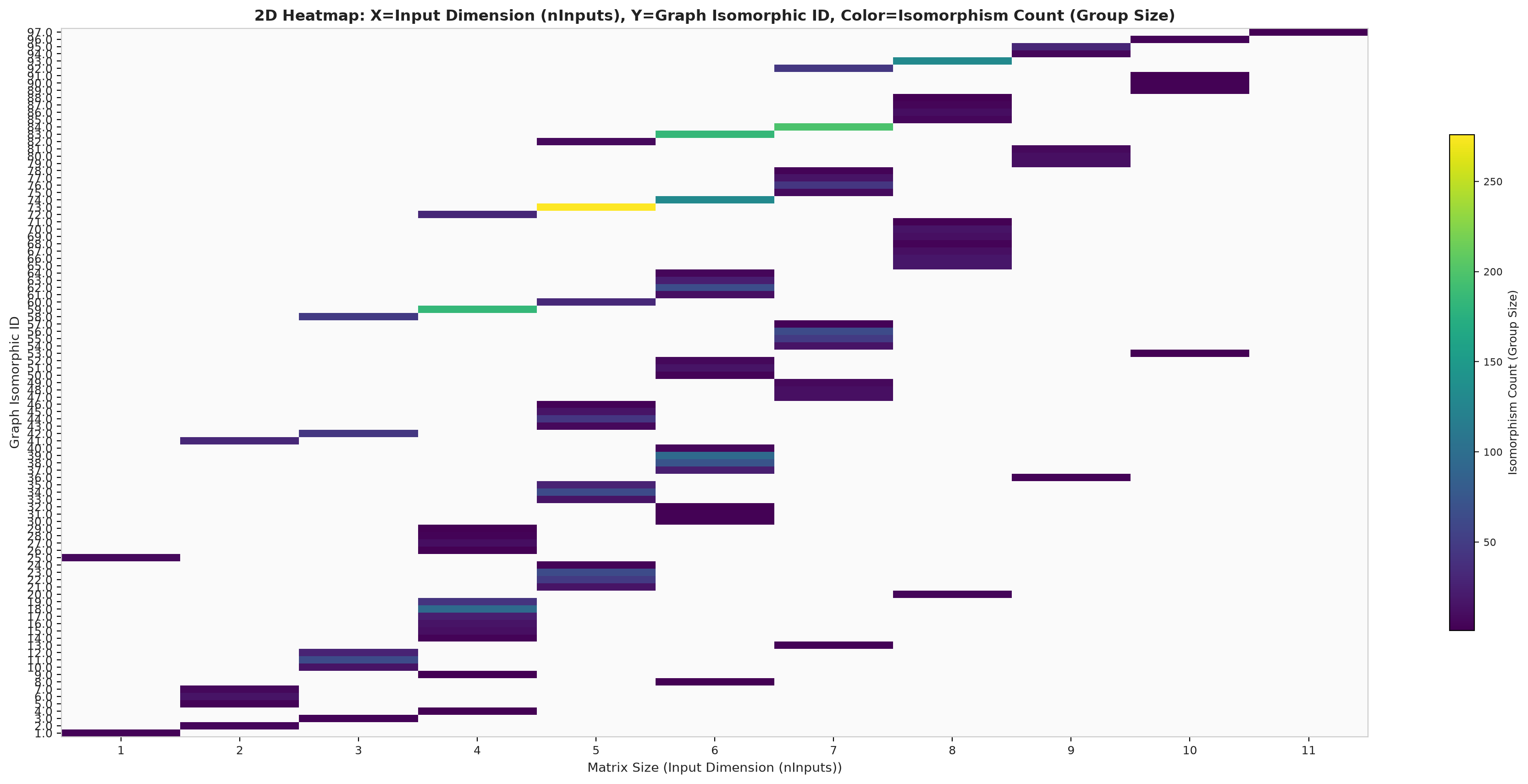}
    \caption{Two-dimensional heatmap showing isomorphism class cardinality as a function of input matrix dimension ($n_{\mathrm{inputs}}$) and class index for matrix $M011$.}
    \label{fig:heatmap_group_size}
\end{figure}

\paragraph{Isomorphism Class Degeneracy}
Figure~\ref{fig:heatmap_group_size} illustrates isomorphism class cardinality across input dimensions 1 through 11 for matrix $M011$ (encompassing 97 distinct classes). While the majority of equivalence classes across both low and high dimensions exhibit small populations (cardinality $< 50$), substantial structural degeneracy is concentrated within intermediate dimensions. Class cardinality peaks sharply at dimension 5 (class index 74, exceeding 250 constituent graphs), with secondary elevated clusters (cardinalities between 150 and 200) observed across dimensions 4, 6, and 7.

\FloatBarrier
\section{Discussion}
\label{sec:discussion}
To synthesize these topological observations and their dependence on input configurations, the mapping can be conceptualized through the lens of an information-routing network. In this framework, the graph vertices correspond to addressable informational states, while directed edges denote admissible computational transitions or information flow between them. The input dimension and state-vector support govern the input bandwidth—the diversity and volume of seed states injected into the matrix transformation simultaneously.

This perspective provides an intuitive explanation for the observed structural transitions:
\begin{itemize}
    \item \textbf{Immediate State Span:} Even minimal, single-dimensional inputs generate graphs with high vertex saturation ($N \approx 8$), indicating that the underlying unitary operations immediately disperse sparse signals across the full accessible state space.
    \item \textbf{Transition Density and Cohesion:} As the input dimension increases, the network experiences a surge in transmission channels (monotonically rising edge counts and self-loops). This densification counteracts structural fragmentation, bridging previously disjoint subgraphs into integrated, strongly connected topologies.
    \item \textbf{Topological Attractors:} The extreme non-uniformity in isomorphism class sizes—where specific equivalence classes exhibit degeneracy counts reaching 56 or exceeding 250—suggests the presence of structural attractors. Rather than generating arbitrary configurations, the matrix transformations preferentially collapse varied input vectors into a conserved set of highly symmetric routing topologies.
    \item \textbf{Feedback and Redundancy:} The discrete, quantized spikes in simple cycle counts (reaching peaks of 2,365 cycles) reflect localized regimes of high structural recurrence. At these specific operational dimensions, the matrix facilitates extensive closed-loop feedback pathways, enabling cyclical state preservation rather than purely feed-forward dissipation.
\end{itemize}

\section{Conclusion}
We'd like to take this opportunity to thank Neil Sloane et.al~\cite{SloaneHadamard} and Wojciech Bruzda et.al \cite{CHM_Catalogue, tadej2006concise} for maintaining a comprehensive collection of Hadamard matrices. Without these databases, this paper is not a possibility. The collection of graph algorithms provided by SageMath~\cite{sagemath} has been used for analyzing the TSS graphs. \\

Through this analysis we conclude that Hadamard matrices both real and complex tend to have similar TSS graphs with the exception of introducing phases by the latter. Superpositions introduce non-uniform probability spaces and can be useful for directing information flow to other TSS graphs for further computation. Isomorphic graph properties are nearly uniform based on the the no of superposed inputs. Isomorphic property of TSS can be potentially used for \emph{variables declarations} for the assembly language (yet to be designed) for a Quantum Computer. \\ 

It's worth pointing out that the no of TSS subgraphs turn out to be very large and there's indeed computational burden in processing Hadamard matrices beyond order $16$ ($4$ qubits). So, we've restricted the analysis to order $8$. \\

The analysis is an extension of our previous TSS paper~\cite{lewis2026} illustrating their utility in building Quantum Algorithms paving way for developing \emph{abstractions} required to build software for Quantum Computers. \\

\nocite{*} 
\printbibliography
\appendix
\section{AppendixA}
\subsection{Relation of Isomorphism and the Input}
In \ref{sec:results} we mentioned that the different numbers of the input produce different isomorphism classes, here we present 2 such examples to prove our point.
\begin{figure}[H]
    \centering
    \includegraphics[width=0.95\linewidth]{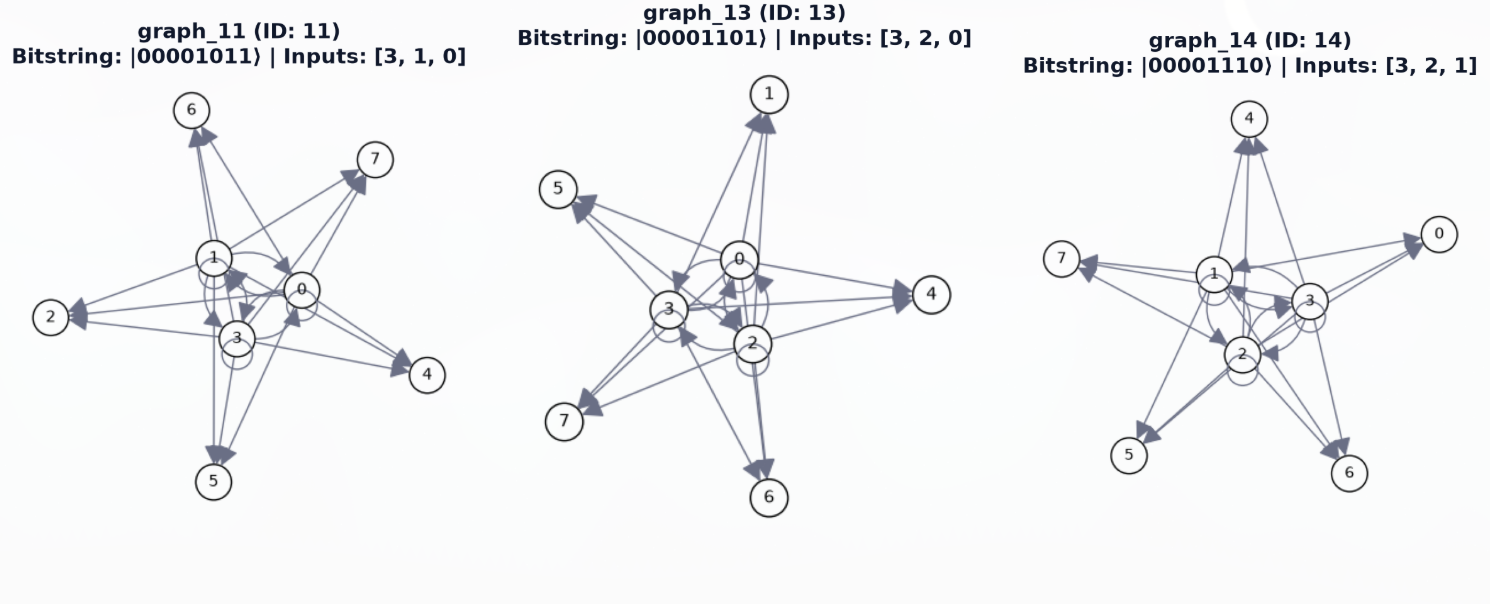}
    \caption{Isomorphic graphs generated for the same number of inputs}
    \label{fig:iso_1}
\end{figure}
\begin{figure}[H]
    \centering
    \includegraphics[width=0.95\linewidth]{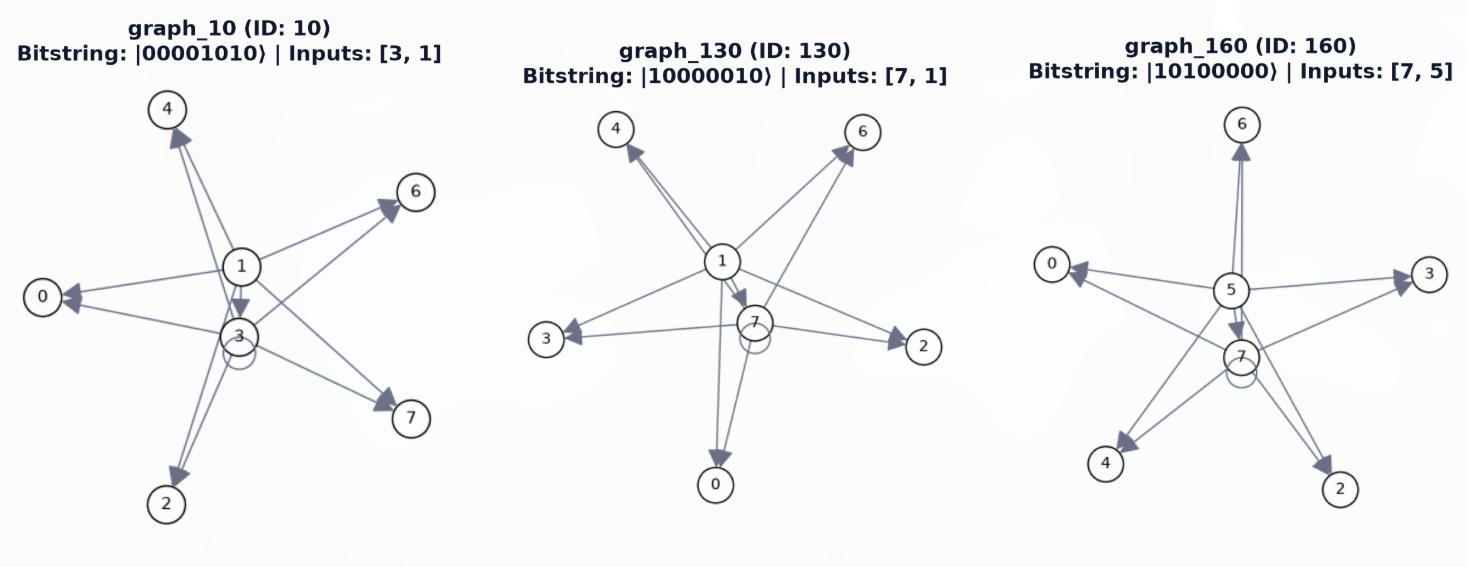}
    \caption{Isomorphic graphs generated for the same number of inputs}
    \label{fig:iso_2}
\end{figure}
\FloatBarrier
The bitsrings in the figures \ref{fig:iso_1} and \ref{fig:iso_2} refer to the superposition of different states in the input but they generate graphs that are isomorphic.

\end{document}